\documentclass{iau}
\usepackage{graphicx}
\usepackage{afterpage}

\usepackage[colorlinks=true,linkcolor=blue,citecolor=blue,urlcolor=blue]{hyperref}

\title[Center-to-limb variation 
on full-disk observations] 
{Center-to-limb variation of spectral lines and their effect
on full-disk observations}

\author[Alexander G.M. Pietrow \& Adur Pastor Yabar]   
{Alexander G.M. Pietrow$^1$
 \and Adur Pastor Yabar$^2$}

\affiliation{$^1$Leibniz-Institut für Astrophysik Potsdam (AIP), An der Sternwarte 16, 14482 Potsdam, Germany \\ email: {\tt apietrow@aip.de} \\[\affilskip]
$^2$Institute for Solar Physics, Dept. of Astronomy, Stockholm University, Albanova University Centre, SE-106 91 Stockholm, Sweden}

\pubyear{2008}
\volume{xxx}  
\pagerange{119--126}
\jname{Title of your IAU Symposium}
\editors{A.C. Editor, B.D. Editor \& C.E. Editor, eds.}
\begin{document}

\maketitle

\begin{abstract}
An accurate description of the center-to-limb variation (CLV) of stellar spectra is becoming an increasingly critical factor in both stellar and exoplanet characterization. In particular, the CLV of spectral lines is extremely challenging as its characterization requires highly detailed knowledge of the stellar physical conditions. To this end, we present the Numerical Empirical Sun-as-a-Star Integrator (NESSI) as a tool for translating high-resolution solar observations of a partial field of view into disk-integrated spectra that can be used to test common assumptions in stellar physics.  
\end{abstract}

\firstsection 
\section{Introduction}

As telescopes and instruments provide ever more detailed observations of the night sky, previously negligible effects come to the forefront. This in turn requires once simple models to incorporate increasing complexity to explain the observed data, often with no clear way to test their appropriateness. One such property is the CLV of stellar spectra, which, along with its associated parameterisation, has become increasingly important for the characterisation of stars and exoplanets. For example, incorrect limb darkening curves in transit observations lead to an overestimation of the planet radius \cite[(e.g. Mandel \& Agol 2002, and Chakraborty \etal\ (submitted), Canocchi \etal\ (submitted))]{}.
Currently, most such studies rely on continuum limb darkening curves obtained from models such as the PHOENIX \cite[(Husser \etal\ 2013)]{} or ATLAS \cite[(Kurucz 1970)]{} codes, which ignore the CLV of spectral lines. 

While \cite[Maxted (2023)]{} has shown that it is a valid approximation to use continuum CLV for broadband ( $\approx 500$ nm) observations, this is not necessarily the case when studying spectra \cite[(e.g. Dineva \etal\ 2020, Reiners \etal\ 2023, and Pietrow \etal\ 2023)]{}. In addition, substructures on the disk such as sunspots and plage can also affect transits and radial velocity measurements \cite[(e.g. Meunier et al. 2010).]. For this reason,  active regions must also be modelled together with their CLV, which have been shown to differ from that of the quiet Sun \cite[(e.g. Oranje 1983 and Cretignier et al. 2023)]{}. Therefore, theoretical models require further development so that they can  account for this CLV behavior.

The Sun is an obvious test bed for such studies, as many small-scale features on its surface can be resolved due to its large angular size and proximity to Earth. However, most high spatial and temporal resolution observations are inherently limited by a small field of view, making it difficult to compare them in a Sun-as-a-star setting \cite[(Otsu \etal\ 2023 and Pietrow \etal\ 2023c)]{}. The NESSI code\footnote{\url{https://github.com/apy-github/NESSI}}, which is publicly available,  has been developed to overcome this limitation and translate such observations into disk-integrated spectra. This allows specific aspects to be studied in isolation from other events on the solar disk.  

\section{Numerical Empirical Sun-as-a-Star Integrator }
\begin{figure}[b]
\begin{center}
 \includegraphics[width=\textwidth]{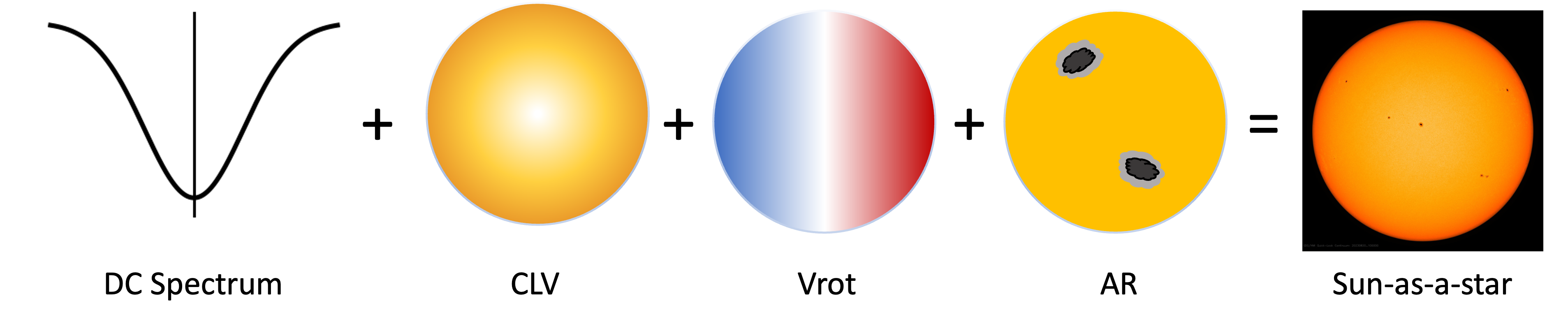} 
 \caption{A cartoon showing the ingredients required for creating a disk-integrated or Sun-as-a-star profile.}
   \label{fig1}
\end{center}
\end{figure}

In its base function, NESSI works in the same way as other stellar integrator codes such as the Spot Oscillation And Planet code \cite[(SOAP, Boisse \etal\ 2012, Dumusque \etal\ 2014, Zhao \& Dumusque 2023)]{Dumusque14,Zhao23}, except that it works with empirical inputs from telescopes such as the Swedish 1-m Solar Telescope \cite[(SST, Scharmer \etal\ 2003)]{} and the Interface Region Spectrograph \cite[(IRIS, De Pontieu \etal\ 2014)]{}. Following the recipe shown in Fig. \ref{fig1}, the code requires a spectral profile or wavelength range of the quiet Sun (QS) at disk center (DC), along with a set of CLV curves for each wavelength point. It also requires a differential rotation velocity profile and the orientation of the rotation axis with respect to the observer, so that it can evaluate the spectra of each point on the solar disk and integrate them into a single SAAS profile. In addition, it is possible to evaluate the effect of small scale features on the integrated spectrum by replacing some grid points with these features CLV (sunspots, plage, flares, etc.).

\section{Quiet Sun simulations}

In order to test the method, we simulated two Sun-as-a-star (SAAS) line profiles for H$\alpha$ and O\,{\sc i}~7772~{\AA}. Both lines need to be modeled in 3D NLTE \cite[(Leenaarts \etal\ 2012, and Bergemann \etal\ 2021)], and therefore will typically not be properly recreated by stellar synthesis codes that focus on producing well-sampled spectral lines. However, since all relevant physical effects are captured in the CLV observations, this method should approximate the final integrated profiles with high accuracy.

We used the CLV from the Institut für Astrophysik Göttingen (IAG) spectral atlas \cite[(Ellwarth \etal\ 2023)]{} and SST line CLVs \cite[(Pietrow 2023b)]{}, respectively. In both cases, we used the $\rm \mu$=1 line profile as the DC profile, the differential rotation described by \cite[ Balthasar \etal\ (1986)]{Balthasar86} and assumed a quiet Sun disc without any other feature. These lines are then compared to continuum limbdarkening curves based on \cite[Neckel \& Labs (1994)]{}, and the full-disk IAG flux atlas \cite[(Reiners \etal\ 2016)]{}, which in case of the oxygen line is convolved down to the resolution of the SST data using the STIC python library \cite[(de la Cruz Rodr{\'\i}guez \etal\ 2019)]{}.

\begin{figure}[p]  
    \centering
    \includegraphics[width=1\textwidth]{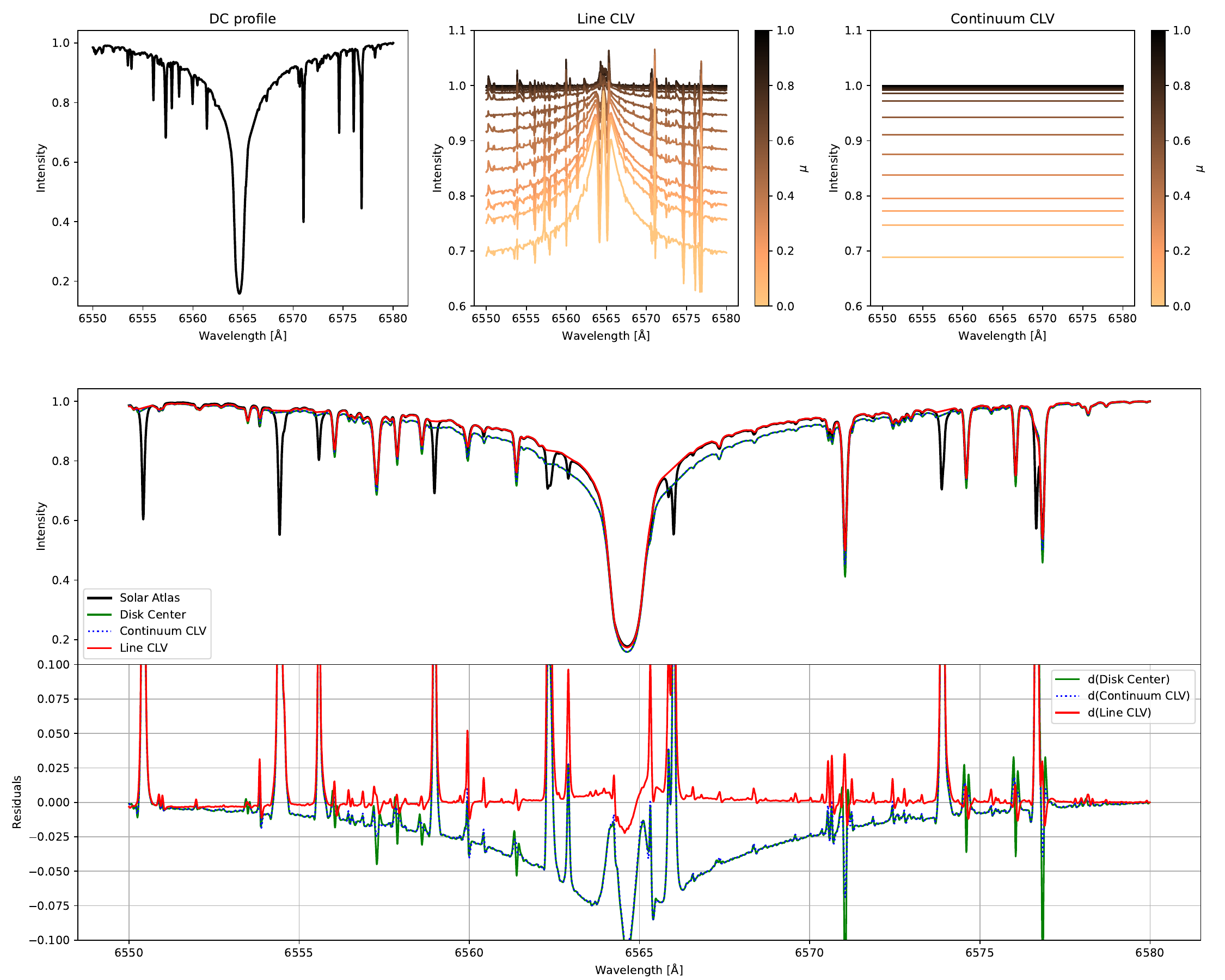} 
    \caption{A Sun-as-a-star H$\alpha$ profile of the QS compared with the IAG flux atlas. \textbf{Top Left:} A disk center profile. \textbf{Top Middle:} Line CLV curves. \textbf{Top Right:} Continuum CLV curves. \textbf{Middle:} A comparative plot showing the atlas, and three synthesized Sun-as-a-star profiles. \textbf{Bottom:} Difference between each profile and the atlas reference.}
    \label{fig2}

    \vspace{1cm}  

    \includegraphics[width=1\textwidth]{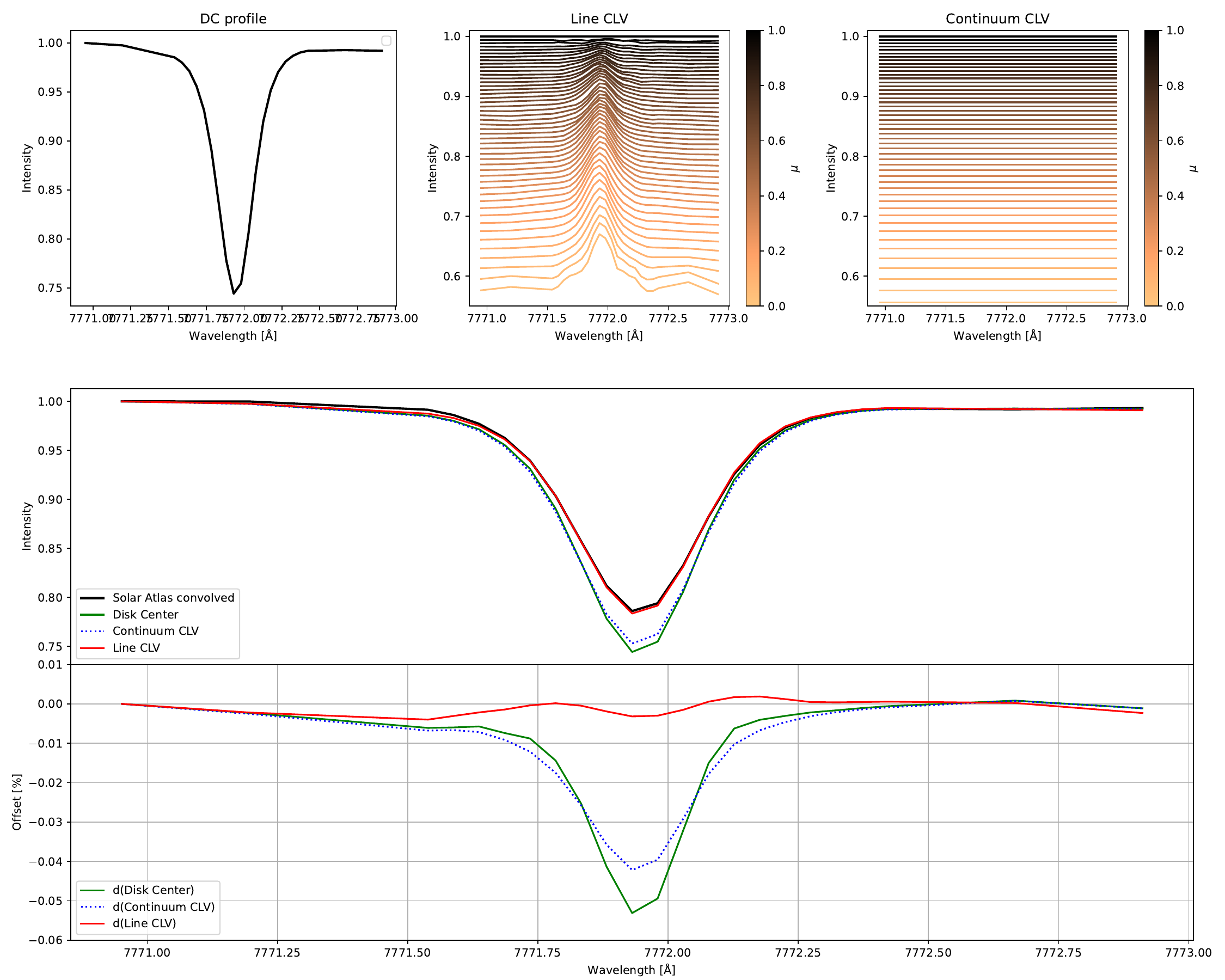} 
    \caption{Same as for Fig. \ref{fig2}, but for the O\,{\sc i}~7772~\AA\ line taken with a lower spectral resolution.}
    \label{fig3}
\end{figure}

The results of both experiments are shown in Figs. \ref{fig2} and \ref{fig3}, where the difference between line CLV and more traditional limb-darkening curves is shown. SAAS profiles are synthesized for both cases, which are then compared with the spectral atlas and a disk center profile, which is equivalent to assuming no limb darkening at all. Evidently, the profile constructed using the line CLV results in a much better fit in both cases, as the residuals are on the order of one percent. The larger peaks in the residual spectrum are at the locations of telluric lines that were removed in the IAG flux atlas. The gaps in these areas have been interpolated, resulting in large offsets from the atlas spectrum. Any remaining difference between this profile and the reference is likely due to the fact that the IAG flux atlas is averaged over several weeks worth of non-quiet disk observations, resulting in an unknown contribution from surface active regions. The effect of this will be investigated in the future in a comparative time-resolved study with high-resolution Sun-as-a-star telescopes such as those on HARPS-N \cite[(Dumusque \etal\ 2015)]{} and PEPSI \cite[(Strassmeier \etal\ 2018)]{}, where the size and location of the active regions are known.   

In summary, NESSI incorporates empirical line CLV into its SAAS integration, resulting in a better agreement with full-disk spectral observations compared to using continuum model CLV only. Thus, if provided with empirical/theoretical CLV for small-scale features, it facilitates a better evaluation of the impact these have over the integrated spectrum. The NESSI code is publicly available on the project github (see link on P1).
\\\\
\textbf{Acknowledgements}\\
\scriptsize{We thank Dominique Petit dit de la Roche, and Malcolm Druett for their stimulating discussion on the topic. Co-funded by the European Union (ERC, MAGHEAT, 31004331). Views and opinions expressed are however those of the authors only and do not necessarily reflect those of the European Union or the European Research Council. Neither the European Union nor the granting authority can be held responsible for them. The Institute for Solar Physics is supported by a grant for research infrastructures of national importance from the Swedish Research Council (registration number 2017-00625).}
 
\end{document}